\documentclass{aa}

\usepackage{epsfig}

\newcommand{\msun}{$M_{\odot}$}

\newcommand{\rsun}{$R_{\odot}$}

\newcommand{\mk}{$M_{\rm K}$}

\newcommand{\ik}{$I-K$}
\newcommand{\vk}{$V-K$}

\newcommand{\od}{$\bigcirc \hspace*{-1.7ex}\bullet$\,}
\newcommand{\cross}{{\large +}\,}

\def\msol{\mbox{M}_\odot}
\def\lesssim{\mathrel{\hbox{\rlap{\hbox{\lower4pt\hbox{$\sim$}}}\hbox{$<$}}}}
\def\gtrsim{\mathrel{\hbox{\rlap{\hbox{\lower4pt\hbox{$\sim$}}}\hbox{$>$}}}}

\begin{document}

\thesaurus{08(01.1, 02.1, 05.3, 06.3, 12.1, 14.2)}
\title{Are the red dwarfs in cataclysmic variables main-sequence stars?} 
\author{K. Beuermann
\inst{1}
\and I. Baraffe
\inst{2}
\and U. Kolb
\inst{3}
\and M. Weichhold
\inst{1}
}

\offprints{beuermann@uni-sw.gwdg.de}
\institute{Universit\"ats-Sternwarte, Geismarlandstr. 11, D-37083 G\"ottingen,
Germany 
\and C.R.A.L. (UMR 5574 CNRS), Ecole 
Normale Sup\'{e}rieure de Lyon, F-69364 Lyon Cedex 0.7, France
\and Astronomy Group, University of Leicester, University Road, Leicester LE1 7RH,
United Kingdom}
\date{Received June 29, 1998 / Accepted September .. , 1998} 

\authorrunning{K. Beuermann et al.}
\titlerunning{Secondaries in CVs}
\maketitle


\begin{abstract}
We show that the secondaries in short-period cataclysmic variables
with orbital periods $P\,< 3$\,hr are close to the solar-abundance
main sequence defined by single field stars.  In cataclysmic variables
with $P\,> 3$\,hr, the earliest spectral types at a given period
correspond to main sequence stars, while the majority of secondaries
have later spectral types. Possible causes are nuclear evolution prior
to mass transfer and lack of thermal equilibrium due to mass
transfer. A comparison with evolutionary sequences obtained with
up--to--date stellar models implies unusually high transfer rates and
a large fraction of systems with evolved donors. There is no evidence
for a secondary of low metallicity in any of the well-studied
cataclysmic variables.

\end{abstract} 

\section{Introduction} 

Fifteen years ago, Echeverr\'{\i}a (1983) addressed the question
whether the secondaries in cataclysmic variables (CVs) are main
sequence (MS) stars. While he concluded that they have, in general,
later spectral types than MS stars of the same mass, his study was
limited by poor statistics, particularly below the period
gap. Patterson (1984) concluded that the empirical zero-age main
sequence (ZAMS) adequately described the secondary stars except for
CVs with orbital periods $P\,\ga8$\,hr. Similarly, Smith \& Dhillon
(1998) focus on systems with estimated secondary masses and radii and
conclude that CV secondaries with $P\,< 8$\,hr are, as a group,
indistinguishable from MS stars in detached binaries. On the other
hand, Friend et al.\ (1990) found that the secondaries in a
substantial number of CVs at shorter periods are too cool to pass
credibly for ZAMS stars. In this paper, we point out that secondaries
in CVs deviate noticeably from field MS stars for a certain range in
orbital period and discuss these deviations in the framework of
evolutionary models.

Roche geometry and Kepler's laws define the orbital period $P$ of a CV
which harbours a secondary of mass $M_2$ and radius $R_2$  as 
\begin{equation}
P =
\left(\frac{R_2/R_{\odot}}{0.234\,f(q)}\right)^{3/2}(M_2/M_{\odot})^{-1/2}
\qquad {\rm hr}
\end{equation}
\noindent
where $f(q)$ varies between 1.032 and 0.990 for mass ratios $q =
M_2/M_1 \le 1$, and $M_1$ and $M_2$ are the masses of white dwarf and
secondary star, respectively. For the secondaries in most CVs, $R_2$
and $M_2$ are ill-determined and the only well-determined quantity is
the spectral type $Sp$.  Therefore the $Sp-P$ diagram of CVs is an
excellent observational tool to study properties of CV
secondaries. Unlike Smith \& Dhillon (1998), we focus on this diagram.

In order to compare the secondaries in CVs with MS stars, we need an
equivalent theoretical $Sp(P)$ relationship for field stars, i.e. the
period $P$ of a CV in which a given field star as a secondary would
just fill its Roche lobe. In the present paper, we derive this
relation, making use of the recent convergence of the theoretical and
observational descriptions of the lower MS.

On the theoretical side, significant progress towards an accurate
description of the mechanical and thermal properties of low-mass stars
has been made by the use of improved internal physics and outer
boundary conditions based on non-grey atmosphere models. Evolutionary
calculations based on the interior models of Chabrier and Baraffe
(1997), combined with recent {\it NextGen} atmosphere models and
synthetic spectra of Hauschildt et al. (1998; see also Allard et
al. 1997), have led to a much improved representation of the observed
properties of M-dwarfs (Baraffe et al. 1995, 1997, 1998; henceforth
summarized as BCAH).  These models provide mass-colour and
mass-magnitude relationships which can be directly compared to
observed quantities. On the observational side, the application of the
{\it NextGen} models to the analysis of low-resolution optical/IR
spectra has improved to the point that an acceptable radius and
temperature scale is in view (Leggett et al. 1996, henceforth L96).

\section{The lower main sequence}

Chabrier \& Baraffe (1995) and Baraffe \& Chabrier (1996) showed that the
BCAH models closely reproduce the observed radii of the visual
binaries CM Dra and YY Gem, and the mass-spectral-class relationship
of very-low-mass stars. More recently, the models have been shown to
reproduce the lower MS of globular clusters (BCAH97), observed
mass-magnitude relationships in the V- and K-bands, and
colour-magnitude diagrams in near-infrared colors (BCAH98). A detailed
description of the input physics for the most recent low-mass star
models used in our study can be found in Baraffe et al. (1998; see
also references therein).

Figure~1 compares stellar radii $R/$\rsun~ determined
quasi-observationally by L96 with model radii calculated by BCAH. This
comparison is made on a luminosity scale with luminosity represented
by the absolute magnitude \mk~in the K-band.  \mk~is an
observationally well-determined quantity for most nearby field stars
and is well-reproduced by the theoretical models (see BCAH98). The
solid curve represents the ZAMS for solar metallicity [M/H] = 0. The
dotted curve is the 0.1 Gyr isochrone for [M/H] = 0. At this age, the
less massive ($M \lesssim 0.4 \msol$) stars are still in their pre-MS
contraction phase. At a given \mk, decreasing metallicity implies a
reduction in radius by some 8\% per dex in [M/H] ([M/H] = $-0.5$,
long-dashed curve, [M/H] = $-1.5$, short-dashed curve). The data
points represent the observationally determined radii of eight young
disk stars (YD, $\bullet$), four old disk stars (OD or O/H, $\bigcirc$),
and four halo stars (H, $\bigtriangleup$) (L96). Four of these stars
are binaries (Gl65AB, 129, 206, and 268). We assume they have two
identical components and include the mean values in Fig. 1. The 12 YD/OD
stars and one H star have radii very close to those predicted for
[M/H] = 0 which is in line with the lack of a one-to-one relation
between kinematic population class and metallicity (e.g. Leggett 1992,
henceforth L92). The average ratio of the observed over the
theoretical radii for the eight YD stars alone is $1.010 \pm 0.010$,
for the 12 YD/OD stars it is $1.020 \pm 0.010$. On the average, the
radii of these stars with {\it bona fide} near-solar metallicity agree
with the ZAMS model radii for stars of the same luminosity within
2\%. Individual stars deviate by up to 6\% in both directions, but
they are all within 2 standard deviations of the [M/H] = 0 model. Two
halo stars have radii as expected for their low metallicity (L96),
confirming the spread in radius as a function of metallicity predicted
by the BCAH models.

\begin{figure}[t] 
\psfig{file=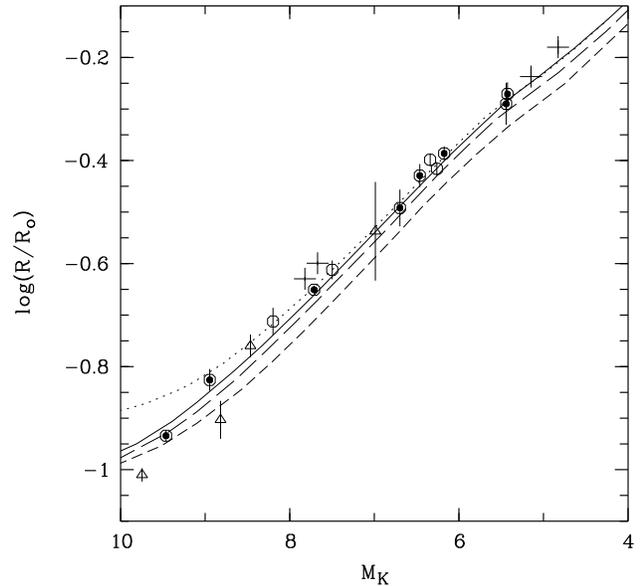,width=8.8cm}
\caption[]{Radii of low-mass main-sequence stars
as a function of absolute magnitude \mk~in the K-band. Observed points
are from Leggett et al. (1996, L96) for YD (\od), OD
($\bigcirc$), and H stars ($\bigtriangleup$). Crosses (\cross) indicate the
individual components of the late-type binaries CM Dra and YY Gem. The
error bars include the uncertainties in the parallaxes. The
theoretical curves are from Baraffe et al. (1997, 1998) for the ZAMS
at metallicity [M/H]\,= 0 (solid), and for an age of 10 Gyr at
metallicities [M/H]\,= $-0.5$ (long dashes) and [M/H]\,= $-1.5$ (short
dashes). The dotted curve is for [M/H]\,= 0 and an age of 0.1 Gyr.}
\end{figure}

Also shown (as \cross) are the radii of the binary components of
YY Gem and CM Dra. The luminosity of YY Gem is based on the HIPPARCOS
parallax, $\pi = 63.3$\,mas (Jahreiss, private communication), the one
of CM Dra on the ground-based parallax, $\pi = 69.2$\,mas (van Altena
et al. 1995). The CM Dra points indicate radii larger than expected
from the [M/H] = 0 model by $12-13$\%, a discrepancy which would
disappear for the larger distance suggested by Chabrier \& Baraffe
(1995).  

It is interesting to compare these quasi-observationally derived radii
of dwarf stars with those predicted by the Barnes-Evans relation
(Barnes \& Evans 1976) which is almost entirely based on giants and
supergiants, but thought to be applicable to dwarfs, too (Lacy
1977). For nine stars in common between Lacy and L96, the average
ratio of the radii (after conversion to the same parallax) is $R({\rm
Lacy})/R({\rm L96}) = 1.09\pm0.05$ with individual values of the ratio
ranging from 0.90 to 1.32. This systematic difference, however, seems
to be present only over a restricted range in spectral type, notably
early M, where the Barnes-Evans relation displays a break due to the
transition between two power-law approximations. We consider the L96
radii to be more appropriate for dwarf stars and further discuss the
difference to the Barnes-Evans radii in a forthcoming paper (Beuermann
\& Weich\-hold 1998).

Clemens et al. (1998) claim that substructure exists in the $R-M$
diagram which is not reproduced by the models. Such substructure is
not evident from the $R($\mk) relationship of Fig.\,1 where it would be
expected at \mk~$\simeq 8$. Moreover, Kolb et al. (1998) have shown
that the hypothetical period distribution of CVs based on the
mass--radius relationship derived by Clemens et al. (1998) does not
match the observed distribution around the period gap.

\begin{figure}[t] 
\psfig{file=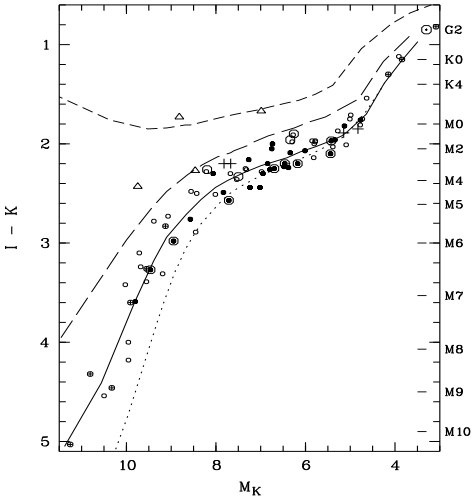, height=11.0cm, width=8.8cm}
\caption[]{ Colour-magnitude diagram (\ik) vs. \mk~for the
sample of main sequence stars defined in Appendix A. Symbols for the
stars from L96, CM Dra, and YY Gem are as in Fig. 1. Other stars: YD
($\bullet$), OD ($\circ$), no kinematic class ($\oplus$), Sun
($\odot$). The curves are as in Fig. 1.}
\end{figure} 

Radii based on analyses with the {\it NextGen} stellar atmosphere
models are available only for the few stars shown in Fig. 1. However,
the Figure allows one to predict the radius of a star of about solar
metallicity and of given luminosity by adopting the observationally
confirmed theoretical $R($\mk) relationship for [M/H] = 0 (solid
line). The mean offset of the corresponding data points from the
theoretical curve by 1--2\% is easily accounted for by one or more of
the following factors: (i) systematic errors in the observationally
derived radii ($\sim 2$\%, L96); (ii) the mean error in the K-band
fluxes used ($<3$\%);
and (iii) the uncertainties in the age and the metallicity of the
observed objects.

In order to define the shape of the MS in sufficient detail, we
supplement the sample of 12 YD/OD stars in L96 by 67 proven or
presumed single YD/OD stars.  This sample is defined in Appendix A\,1
and A\,3 and extends from the Sun down in luminosity to the transition
from the stellar to the brown-dwarf regime (represented by
GD165B). These stars are mostly from the list of single stars of Henry
\& McCarthy (1993, their Table 3) which are selected on the base of
speckle interferometry. It is important to exclude unrecognized
binaries since they would be mistaken for single stars of higher mass
and radius and would blur the empirical MS as well as the
corresponding $Sp-P$ diagram. Since we want to concentrate on stars of
solar metallicity, we have included no further H stars beyond the
three in the L96 subsample.

\begin{figure}[t] 
\psfig{file=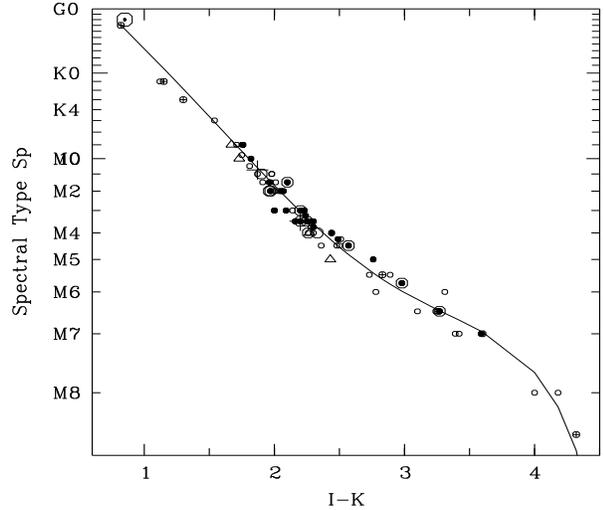, height=11.0cm, width=8.8cm}
\caption[]{ $Sp$ vs. $I-K$ diagram for the stars of Fig.~2.
Crosses (\cross) indicate the mean components of the binaries CM~Dra
and YY~Gem.  The Sun is included at $Sp = $G2. The solid curve is the
least-squares fit to the YD and OD stars as given by Eq. (2).}
\end{figure}  

Figure~2 shows the colour-magnitude diagram\footnote{All I magnitudes
refer to the Cousins system, K magnitudes to the CIT system.}
\mbox{(\ik)} --\mk~of the complete sample along with theoretical
curves for the same BCAH models as in Fig. 1.  These models reproduce
the observed (\ik)$-$\mk~ diagram exceedingly well. For fixed age and
colour, they predict a drop by roughly 1 mag in \mk~for a decrease in
[M/H] by $-0.5$ dex. For the observed sample of field M--stars of
mixed age and composition, we expect the bright limit of the
distribution to be populated by ZAMS stars with near-solar metallicity
plus an admixture of pre-MS stars.  For the spectral range $Sp$ =
M0$-$M5, the bright limit is, in fact, dominated by YD stars which
typically have near-solar metallicity (e.g.  L96). The lower part of
the diagram is dominated by stars classified kinematically as OD, but
at least some of these have spectroscopically determined [M/H] in the
range of $-0.5$ to $+0.5$ (e.g. Jones et al. 1996, Schweitzer et
al. 1996) and, hence, are approximately solar-like, too. The lack of
stars falling near the 0.1\,Gyr isochrone (dotted curve) indicates the
scarcity of low-mass pre-MS stars in the sample.

The conversion from \ik~ to $Sp$ used for the theoretical models (and
indicated on the right-hand side of Fig. 2) is based on the $Sp($\ik)
relation for the YD/OD stars of our sample which is shown in
Fig. 3 (solid curve). It is represented by the third-order polynomial 
\begin{equation}
X=48.93-36.94(I-K)+10.313(I-K)^2-0.998(I-K)^3
\end{equation} 
where the spectral types of M-, K-, and G-stars are given as a
function of \ik~ by $Sp = $M$(10-X)$ for \mbox{$X\le10$}, $Sp =
$K$(18-X)$ for $10 < X \le 18$, and $Sp = $G$(28-X)$ for $18 < X \le
28$, respectively. For M-stars, this relation is very close to those
of L96 and Kirkpatrick \& McCarthy (1994). For stars earlier than K3,
the relation may be less accurate and off by 1--2 subclasses.

\section{The $Sp-P$ diagram for main--sequence field stars}

\begin{figure}[t] 
\psfig{file=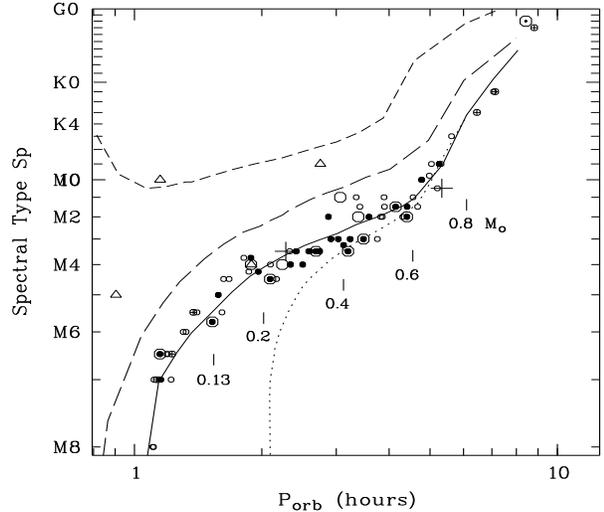, height=11.0cm, width=8.8cm} 
\caption[]{ $Sp-P$ diagram for the single field stars of
Fig. 2.  Crosses (\cross) indicate the mean components of the binaries CM
Dra and YY Gem. The Sun is included at $Sp =$ G2. Theoretical curves
are as in Fig. 1. }
\end{figure} 

\begin{figure}[t] 
\psfig{file=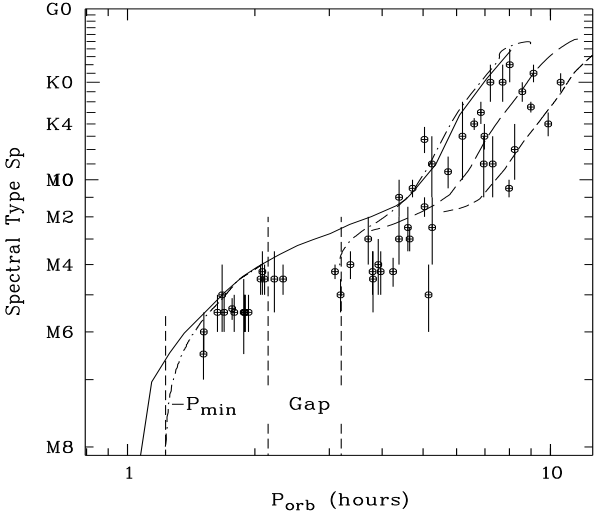, height=11.0cm, width=8.8cm}
\caption[]{ $Sp-P$ diagram for CVs secondary stars. Data
points are from Ritter \& Kolb (1998) (see also Appendix B). The solid
curve indicates the location of ZAMS stars with [M/H]=0 from
Fig. 4. In addition, evolutionary sequences for an initial
1\,\msun~secondary with constant mass loss rate are displayed for
different starting conditions: ZAMS model for [M/H]=0 (dot-dashed
curve), evolved models with reduced central hydrogen abundance (dashed
curves, see text). Note the lack of observed secondaries of low
metallicity which would populate the upper left of the diagram. }
\end{figure} 

The essential observational information on the main-sequence nature of
the secondaries in CVs is contained in the $Sp-P$ diagram. It is
important, therefore, to determine the locus of main--sequence field
stars of different metallicity in such a diagram. For this purpose, we
transform the data points and the BCAH model curves of Fig. 2 into
the $Sp-P$ plane, using Eq.(1). The results are shown in Fig. 4. For
the theoretical curves, the periods $P$ are readily determined from
the model values for $R$ and $M$, while the spectral types $Sp$ are
derived from the model values of \ik. As the latter may still be in
error by up to $0.1-0.2$\,mag for low mass stars, the error in the
inferred $Sp$ can reach 1/2 spectral class. For the observed stars, on
the other hand, $Sp$ is known, but $R$ and $M$ are generally not, with
the exception of the binary components of CM Dra and YY Gem (Metcalfe
et al. 1997; Leung \& Schneider 1978). Quasi-observational values of
$R$ are available for the L96 stars.  For the remaining stars, we use
the BCAH~ZAMS model radii for solar abundances as a function of \mk~
(Fig. 1, solid curve). Except for low-mass pre-MS stars and for some
of the OD stars the implied error in $R$ is $\la 5$\,\%. Masses are
derived for all stars except CM Dra and YY Gem from the appropriate
theoretical $M($\mk) relationship, i.e. the [M/H] = 0 model for the
L96 YD stars as well as the additional YD/OD stars, and the [M/H] =
$-0.5$ and $-1.5$ models for the L96 OD stars and H stars,
respectively. This is an acceptable approach since the $M($\mk)
relationship is in good agreement with the Henry and McCarthy (1993)
binary data (cf. BCAH98). Note that the clustering of the data points
near the model curve for ZAMS stars with [M/H] = 0 is to some extent
artificial as a result of using the corresponding model values for $R$
and $M$.

The model curves in Fig. 4 indicate the fundamental dependencies
expected in the $Sp(P)$ diagram. At a given period, MS stars with
near-solar metallicities have the latest spectral types, whereas stars
of lower metallicity display earlier spectral types. The effect is
clearly noticeable already for a few 0.1\,dex from solar metallicity.
For solar abundances, the model predicts \ik~= 2.40 at $P = 2$\,hr
which corresponds to $Sp \simeq\,$M4. The brightest observed MS stars
at this period have M4.5. The lower boundary to the distribution of MS
stars in Fig. 4 is delineated by the L96 YD stars (encircled filled
circles) and corresponds to the bright limit in Fig. 2.  OD stars
(open circles) and H stars (triangles) are seen to extend into the
upper left part of the diagram.
 
As $P_{\rm min} \simeq 75$\,min is approached, $Sp$ drops rapidly, as
expected from the severe drop in the mass-luminosity relationship
below $\simeq 0.1 \msol$ ($Sp \simeq\,$M6) (cf.  Baraffe and Chabrier
1996). The secondaries in real CVs become degenerate near $P_{\rm
min}$ (e.g.\ Paczy\'nski \& Sienkiewicz 1981). They never reach radii
below $\sim0.1$\,\rsun~ which implies periods somewhat longer than
predicted for MS stars of the same mass. Hence, in real CVs, the curve
would drop even more rapidly reaching very late spectral types already
near 80 min (see Sect. 4 below). Therefore, it is not surprising that
only extremely few secondaries in CVs with $P < 90$\,min have so far
been detected spectroscopically (Howell et al.  1998).

\section{The $Sp-P$ diagram for CVs}

Ritter \& Kolb (1998) have compiled the spectral types of 60 CVs with
orbital period $P < 0.5$\,d. We have checked all original references
and accept 50 classifications which are supported by optical/IR
spectroscopy; we exclude classifications based on photometry only and
a few which appear less compelling. Two classifications from more
recent papers are added. See Appendix B for details and for a list of
the adopted spectral types. The resulting $Sp-P$ diagram is shown in
Fig. 5. Also shown is the location of ZAMS stars with [M/H]=0 from
Fig. 4 (solid curve).
No secondary star is found in the upper left ``halo''part of Fig. 5,
with two possible exceptions noted in the Appendix.  With this caveat,
there is currently no evidence for the existence of halo type CVs with
metal-poor secondaries. Continuing the search for such stars is
clearly important.

>From a comparison of Figs.~4 and 5 we conclude that below the period
gap ($P\la2$~hr) secondaries in CVs are indistinguishable from ZAMS
stars within the observational and theoretical uncertainties, while
above the gap they are mostly cooler than ZAMS stars with the same
mean mass density.

In the following we explore to what extent this discrepancy can be
accounted for by the fact that the secondary transfers mass to the
white dwarf. As a consequence of this mass loss, the secondary
deviates from thermal equilibrium. The deviation is large if the mass
loss time scale is short compared to the star's thermal time
scale. The non--equilibrium reveals itself as a radius expansion for
predominantly convective donors ($M_2\la0.6$~\msun) or contraction for
more massive donors ($M_2\ga0.6$~\msun) compared to their respective
ZAMS radii (e.g.\ Whyte \& Eggleton 1980; Stehle et al.\ 1996). In
contrast, the effective temperature at a given secondary mass is
fairly insensitive to mass loss (e.g.\ King \& Kolb 1998). Hence
donors that have been subjected to mass loss are under--massive and
cooler ($M_2\la0.6$~\msun) or over--massive and slightly hotter
($M_2\ga0.6$~\msun) compared to hypothetical donors with no previous
mass loss in CVs with the same orbital period. This could explain the
late spectral types seen in CVs with $P\la 5-6$~hr, but certainly not
those at longer period.  Nuclear evolution of the secondary star prior
to mass transfer offers a natural explanation for the latter. The
nuclear timescale of stars with mass $\ga1$~\msun is short enough that
nuclear burning can significantly deplete the central hydrogen supply
within the age of the Galaxy. Subsequent mass transfer reduces the
secondary mass and mimics a low--mass star that has an equivalent
nuclear age much longer than a Hubble time.

To illustrate this quantitatively we have calculated several
evolutionary sequences with constant mass loss rate, using the same
input physics as for the BCAH models. At turn--on of mass transfer,
the secondary has a mass of $1$~\msun, is either a ZAMS star with
initial central H abundance $X_c=0.70$ (dot-dashed curve, henceforth
``unevolved sequence''), a moderately evolved MS star with $X_c$
reduced to $0.16$ (long dashes), or a star which is just at the end of
core hydrogen burning ($X_c=4 \,10^{-4}$; short dashes). For all three
cases the adopted transfer rate is $1.5 \times 10^{-9}$~\msun
yr$^{-1}$. Along the last two sequences, henceforth ``evolved
sequences'', the surface H abundance $X_s$ decreases as the convective
envelope reaches H--depleted regions deeper inside the star. Both the
outer boundary conditions of the stellar models and the derived
colours rely on atmosphere models for solar composition (X=0.70). As
this inconsistency makes the models unreliable for $X_s<0.65$ we
stopped the evolved sequences once $X_s$ dropped below this value (at
a mass $0.30$~\msun and $0.34$~\msun, respectively). The unevolved
sequence extends further down to $P=3.2$~hr where the secondary
becomes fully convective (at a mass $0.21$~\msun). In line with the
standard model for the period gap (see e.g.\ King 1988, Kolb 1996, for
reviews) mass loss was terminated at this point and the star allowed
to shrink back to its equilibrium radius. Then mass loss resumed (at
the now shorter orbital period $P=2.1$~hr) with a rate $5 \times
10^{-11}$~\msun yr$^{-1}$, typical for mass transfer driven by
gravitational wave emission. The mass loss rate above the period gap
was chosen such that the sequence reproduces the observed width and
location of the gap, i.e.\ the secondary's radius was larger by a
factor $(3.2/2.1)^{2/3} = 1.32$ than its equilibrium radius when it
became fully convective. The true secular mean transfer rate at longer
orbital periods is not known. Semi--empirical estimates for the
braking rate from a magnetic stellar wind typically give values in the
range $10^{-9} - 10^{-8}$ \msun yr$^{-1}$, consistent with
observational estimates (e.g.\ Warner 1995).

The observed spectral types on both sides of the gap are about M4.5,
supporting the conventional explanation of the gap which predicts that
the secondary masses in the majority of systems just above and below
the gap are the same. Below the period gap, mass loss is so slow that
the star should stay close to the ZAMS, as is, in fact, observed. This
holds until the secondary approaches the period minimum where 
the internal structure is increasingly dominated by electron degeneracy,
H-burning turns off, and the star becomes a brown dwarf.

At longer orbital periods the unevolved sequence fails to give
secondaries cool enough to match observed secondaries, except
immediately above the gap (cf. Fig. 5). The observed spread in $Sp$
for $P\la 5-6$~hr seems to imply much higher transfer rates which
would expand the secondaries further over the equilibrium radius.
Further model calculations demonstrated that for transfer rates in
excess of $5\,10^{-9}$~\msun yr$^{-1}$ the secondary can reach the
region near $P \sim 4$\,hr and $Sp \sim $\,M5. A forthcoming paper
(Kolb \& Baraffe, in preparation) will investigate this possibility in
more detail.

The most evolved sequence (short-dashed curve in Fig. 5) nicely
defines an approximate lower envelope for the observed distribution of
CVs with $P>6$~hr in the $Sp-P$ diagram. (Note that sequences from
initially more massive donors, or with different transfer rate, might
fall slightly below the most evolved sequence in Fig.~5). Additional
effects to reconcile theory with observations seem not necessary in
this period range. As the evolved sequences differ from the unevolved
sequence significantly only when the initial $X_c$ is already very
small, the observed location of CV secondaries implies that in a
fairly large fraction of CVs the donor must have been very close to
the end of core hydrogen burning when mass transfer began (see also
Ritter 1994). Such a large fraction seems to be in conflict, however,
with standard models of CV formation (de~Kool 1992, Politano 1996)
which predict that nascent CVs are dominated by systems with
essentially unevolved donors.

\section{Conclusions} 

A re-evaluation of the properties of the lower MS indicates that
secondary stars in short-period CVs lie on the ZAMS for near-solar
metallicity within the uncertainties. In CVs with orbital period $P >
3$\,hr, the majority of the secondaries is cooler than ZAMS field
stars with solar metallicity, indicative of some expansion. Possible
causes are nuclear evolution prior to mass transfer at the longer
periods and lack of thermal equilibrium due to mass loss at the
shorter ones. A comparison with evolutionary sequences suggests that
an unexpectedly large fraction of CVs has an evolved donor, and that
mass transfer rates for $P\ga4$~hr could be much higher than usually
assumed. Of the secondaries spectroscopically identified so far, none
has a metallicity substantially below solar. Secondaries in CVs with
orbital period $P \la 80$\,min will be of spectral type $Sp \ga\,$M8
and very difficult to detect.

\acknowledgements{We thank Hartmut Jahreiss for providing the
HIPPARCOS parallaxes prior to publication and Frederic Hessman, Boris
G\"ansicke, and Klaus Reinsch for comments and discussions. IB thanks
the Universit\"ats-Sternwarte, G\"ottingen, for hospitality and the
APAPE (PROCOPE contract 97151) for travel support. Theoretical
astrophysics research at Leicester is supported by a PPARC Rolling
Grant.}

\appendix
\section{Sample of main sequence stars}

Our sample of MS stars was collected from the three sources listed
below. Denotations are: YD = young disk, OD = old disk or old
disk/halo, H = halo star, B = binary, SB = spectroscopic binary. For
some stars, different spectral classifications are given in the
literature. In these cases, we have chosen the one more appropriate
for the colours \ik~ and \vk. When no spectral class is available, we
adopt the probable classification appropriate for the colours.

\bigskip

\begin{enumerate}

\item The first group contains the Sun and further nine stars
taken mostly from Reid \& Gizis (1997) which delineate the MS at
spectral types G/K.

\smallskip
\begin{flushleft}
\begin{tabular}[h]{l@{\hspace{6mm}}l@{\hspace{6mm}}l@{\hspace{8mm}}l@{\hspace{7mm}}l@{\hspace{7mm}}l}
\noalign{\smallskip}\hline\noalign{\smallskip}
Name & Pop. & $Sp$ & Name & Pop. & $Sp$ \\ 
\noalign{\smallskip}\hline\noalign{\smallskip}
Gl34A  &  --  & G3V   & Gl488   & OD    & M0 \\
Gl68   &  --  & K1V   & Gl673   & OD    & K7V \\
Gl105A &  OD  & K3V   & Gl820A  & OD    & K5V \\
Gl166A &  OD  & K1V   & Gl820B  & OD    & K7V \\
Gl380  & Y/O  & K7V   & &&\\
\noalign{\smallskip}\hline\noalign{\smallskip}
\end{tabular}
\end{flushleft}

\bigskip
\item The second group contains the 16 M-stars from Leggett et al. (1996,
L96) with individually determined radii. These stars serve as
calibrators for the $R($\mk) relationship.

\smallskip
\begin{flushleft}
\begin{tabular}[h]{l@{\hspace{5mm}}l@{\hspace{5mm}}l@{\hspace{8mm}}l@{\hspace
{5mm}}l@{\hspace{5mm}}l}
\noalign{\smallskip}\hline\noalign{\smallskip}
Name & Pop. & $Sp$ & Name & Pop. & $Sp$ \\ 
\noalign{\smallskip}\hline\noalign{\smallskip}
Gl65AB  & YD,B     & M$6-$& Gl411  & OD    & M2 \\
Gl129   & H,SB  & M0    & Gl494  & YD    & M1.5 \\
Gl195A  & YD     & M2   & Gl699  & OD    & M4 \\
Gl206   & YD,SB  & M3.5 & Gl896A & YD    & M3.5\\
Gl213   & OD     & M4   & Gl908  & OD    & M1 \\
Gl268   & YD,SB  & M4.5 & GJ1111 & YD    & M6.5 \\
Gl299   & H     & M4    & LHS343 & H    & K7:\\
Gl388   & YD     & M3   & LHS377 & H    & M5\\ 
\noalign{\smallskip}\hline\noalign{\smallskip}
\end{tabular}
\end{flushleft}

\bigskip \bigskip \bigskip
\item The third group contains 60 stars from the list of ``single''
red dwarfs in Table 3 of Henry \& McCarthy (1993), of which seven are
already contained in the L96 sample. To these we have added the mean
components of the visual binaries CM Dra and YY Gem, the M8.5 star
TVLM\,513-46546 and GD165B which marks the transition to the
brown-dwarf regime at $Sp \ge $M10 (Kirkpatrick et al. 1995). Hence,
this group contains 
57 stars.

\smallskip
\begin{flushleft}
\begin{tabular}[h]{llllll}
\noalign{\smallskip}\hline\noalign{\smallskip}
Name & Pop. & $Sp$ & Name & Pop. & $Sp$ \\ 
\noalign{\smallskip}\hline\noalign{\smallskip}
Gl15A  & OD    & M1.5   & CM Dra  & OD,B  & M3.5 \\
Gl15B  & OD    & M$4-$  & Gl643   & OD    & M4 \\
Gl54.1 & OD    & M4.5   & Gl644C  & OD    & M7 \\
Gl83.1 & OD    & M4.5   & Gl701   & OD    & M2 \\
Gl105B & OD    & M4     & Gl725A  & Y/O  & M3 \\
Gl109  & YD    & M3+    & Gl725B  & Y/O  & M3.5 \\
Gl166C & OD    & M4.5   & Gl729   & Y/O  & M$4-$ \\
Gl205  & OD    & M1.5   & Gl752A  & OD    & M3\\
Gl229  & YD    & M1.5   & Gl752B  & OD    & M8 \\
Gl251  & Y/O  & M3.5    & Gl809   & OD  & M1 \\
Gl273  & OD    & M3.5   & Gl873   & Y/O  & M3.5 \\
YY Gem & YD,B  &M0.5-1  & Gl880   & OD    & M2 \\
Gl300  &      & M4+     & Gl884   & OD    & M$0-$ \\ 
Gl338A & YD    & M0     & Gl905   & OD    & M5.5 \\
Gl393  & Y/O  & M2      & GJ1002  & OD    & M5.5 \\
Gl402  & Y/O  & M4      & GJ1156  & YD    & M5 \\ 
Gl406  & OD    & M6     & GL1245B &     & M5.5 \\
Gl408  & YD    & M3     & LHS191  &      & M6.5 \\
Gl412A & OD    & M1     & LHS292  & OD    & M6.5 \\
Gl412B & OD    & M6     & LHS523  & OD    & M6.5 \\
Gl445  & OD    & M3.5   & LHS2065 &       & M9 \\
Gl447  & OD  & M4+      & LHS2397a& OD    & M8 \\
Gl450  & OD    & M2     & LHS2471 & OD    & M7 \\
Gl514  & OD    & M1     & LHS2924 & OD    & M9 \\
Gl526  & OD    & M1.5   & LHS2930 & Y/O  & M7 \\
Gl555  &       & M4      & LHS3003 &      & M7 \\
Gl581  & Y/O   & M3.5    & TVLM513 &      & M8.5 \\
Gl625  & YD    & M2     & GD165B  &      & M10 \\
Gl628  & YD    & M3.5  &&&\\
\noalign{\smallskip}\hline\noalign{\smallskip}
\end{tabular}
\end{flushleft}
\end{enumerate}

\bigskip \bigskip
\section{Sample of spectroscopically identified secondaries in CVs}
 
The list of Ritter \& Kolb (1998) contains 60 entries for the spectral
classes of secondary stars in CVs with orbital period $P < 12$\,hr. We
exclude six objects because we judge the spectral evidence as not
sufficiently compelling (UU Aql, TX Col, WW Hor, EX Hya, BD Pav, HX
Peg) and four further objects because the cited spectral type is based
on photometry only (TV Col, CW Mon, X Leo, UU Aqr). The published
spectrum of WW Hor (Beuermann et al. 1987) is very noisy and the
spectral type may be earlier than the quoted M6. For EX Hya, our own
unpublished optical/near IR spectroscopy does not reveal the secondary
while our IR photometry shows what seems to be its ellipsoidal
modulation, suggesting that it is of late spectral type. Two
potentially interesting systems are among those excluded: HX Peg
(Ringwald 1994) may be the only CV with an sdK secondary; BD Pav is
quoted as $Sp = $ K$0-4$ (Ritter \& Kolb 1998, unpublished spectrum),
but as K7$-$M0 by Friend et al. (1990). These two systems would be
located above the MS in Fig. 5, in a region otherwise devoid of CV
secondaries. For a few systems we adjusted the spectral types in the
Ritter \& Kolb list based on the available literature: e.g. VV Pup :~
M5 (Liebert et al. 1978), V2301 Oph :~ M5.5 (Silber et al. 1994,
Ferrario et al. 1995), and BT Mon :~ G8 (Smith et al. 1998). We added
BC UMa~ M6.5 (Friend et al. 1990), RX\,J0203+29 :~ M2.5 (Schwarz et
al. 1998), and RX And :~ K5- (Dhillon \& Marsh 1995). The list gives
the name of the CV, the orbital period in hr, and the adopted spectral
type with estimated error.

\footnotesize

\begin{flushleft}
\begin{tabular}[h]{llllll}
\noalign{\smallskip}\hline\noalign{\smallskip}
Name & P(h) & $Sp$ & Name & P(h) & $Sp$ \\ 
\noalign{\smallskip}\hline\noalign{\smallskip}
BC UMa     & 1.512  & M6.5$\pm$0.5& TW Vir   & 4.384 & M3$\pm$1\\
OY Car      &1.515  & M6$\pm$0.5  & SS Aur   & 4.387 & M1$\pm$1\\
BZ UMa      &1.632  & M5.5$\pm$0.5& RXJ0203  & 4.602 & M2.5$\pm$1\\
VV Pup      &1.674  & M5$\pm$1    & DQ Her   & 4.647 & M3$\pm$0.5\\
V834 Cen    &1.692  & M5.5$\pm$0.5& UX UMa   & 4.720 & M0.5$\pm$0.5\\
HT Cas      &1.768  & M5.4$\pm$0.3& RX And   & 5.037 & K5-$\pm$2\\
Z Cha       &1.788  & M5.5$\pm$0.5& EX Dra   & 5.038 & M1.5$\pm$0.5\\
V2301 Oph   &1.883  & M5.5$\pm$1  & AR Cnc   & 5.150 & M5$\pm$1\\
MR Ser      &1.891  & M5.5$\pm$0.5& EY Cyg   & 5.244 & K7$\pm$2\\   
BL Hyi      &1.894  & M5.5$\pm$0.5& CZ Ori   & 5.254 & M2.5$\pm$1.5\\
ST LMi      &1.898  & M5.5$\pm$0.5& AT Cnc   & 5.729 & K8$\pm$1\\
WW Hor      &1.925  & M6          & AH Her   & 6.195 & K5$\pm$3\\
AR UMa      &1.932  & M5.5$\pm$0.5& SS Cyg   & 6.603 & K4$\pm$0.5\\ 
DV UMa      &2.063  & M4.5$\pm$0.5& V426 Oph & 6.847 & K3$\pm$1\\
HU Aqr      &2.084  & M4+$\pm$0.7 & Z Cam    & 6.956 & K7$\pm$2\\
UZ For      &2.109  & M4.5$\pm$0.5& EM Cyg   & 6.982 & K5$\pm$1\\  
UW Pic      &2.224  & M4.5$\pm$1  & AC Cnc   & 7.211 & K0$\pm$2\\
QS Tel      &2.332  & M4.5$\pm$0.5& TT Crt   & 7.303 & K7$\pm$2\\
AM Her      &3.094  & M4+$\pm$0.3 & V363 Aur & 7.710 & K0$\pm$2\\
MV Lyr      &3.190  & M5$\pm$0.5  & V1309 Ori& 7.983 & M0.5$\pm$0.5\\
V1432 Aql   &3.366  & M4$\pm$0.5  & BT Mon   & 8.012 & G8$\pm$2\\
QQ Vul      &3.708  & M3$\pm$1    & CH UMa   & 8.232 & K6$\pm$2\\
IP Peg      &3.797  & M4+$\pm$0.7 & QZ Aur   & 8.580 & K1$\pm$1\\
VY For      &3.806  & M4.5$\pm$1  & RU Peg   & 8.990 & K2.5$\pm$0.5\\
CN Ori      &3.917  & M4$\pm$1    & SY Cnc   & 9.120 & G9$\pm$1\\
DO Dra      &3.969  & M4+$\pm$0.7 & AE Aqr   & 9.880 & K4$\pm$1\\
U Gem       &4.246  & M4+$\pm$0.5 & DX And   &10.572 & K0$\pm$1\\
\noalign{\smallskip}\hline\noalign{\smallskip}
\end{tabular}
\end{flushleft}
\normalsize
\end{document}